\documentclass[authoryear,final,5p,times,twocolumn]{elsarticle}

\usepackage{graphics}
\usepackage{amssymb}
 \usepackage{lineno}

\journal{J. Atmos. Solar-Terrestr. Phys.}

\begin{document}

\begin{frontmatter}



%\title{A synchronization mechanism linking planetary motions, solar activity, auroras  and climate oscillations}

\title{A shared frequency set between the historical mid-latitude aurora records and the global surface temperature}

%\title{An astronomical model for climate change}


\author{Nicola Scafetta $^{1}$}

 \address{$^{1}$ACRIM (Active Cavity Radiometer Solar Irradiance Monitor Lab)  \& Duke University, Durham, NC 27708, USA.}

\begin{abstract}
Herein we show that the historical records of  mid-latitude auroras from 1700 to 1966 present oscillations with periods of about 9, 10-11, 20-21, 30 and 60 years. The same frequencies are found in proxy and instrumental global surface temperature records since 1650 and 1850, respectively and in several planetary and solar records. Thus, the aurora records reveal a physical link between climate change and astronomical oscillations. Likely, there exists a modulation of the cosmic ray flux reaching the Earth and/or of the electric properties of the ionosphere. The latter, in turn, have the potentiality of modulating the global cloud cover that ultimately drives the climate oscillations through albedo oscillations.  In particular, a quasi 60-year large cycle is quite evident since 1650 in all climate and astronomical records herein studied, which also include an historical record of meteorite fall in China from 619 to 1943. These findings support the thesis that climate oscillations have  an astronomical origin.  We show that a harmonic constituent model based on the major astronomical frequencies revealed in the aurora records is able to forecast with a reasonable accuracy the decadal and multidecadal temperature oscillations from 1950 to 2010 using the temperature data before 1950, and vice versa. The existence of a natural 60-year modulation of the global surface temperature induced by astronomical mechanisms, by alone, would imply that at least 60-70\% of the warming observed since 1970 has been naturally induced. Moreover, the climate may stay approximately stable during the next decades because the 60-year cycle has entered in its cooling phase.\\ \\ \textbf{Please, cite this article as:} Scafetta N., 2012. A shared frequency set between the historical mid-latitude aurora records and the global surface temperature. Journal of Atmospheric and Solar-Terrestrial Physics 74, 145-163.
DOI: 10.1016/j.jastp.2011.10.013. http://dx.doi.org/10.1016/j.jastp.2011.10.013 \\
\end{abstract}


\end{frontmatter}

%\linenumbers

\section{Introduction}

Since ancient times people have claimed that  climate and weather changes are related to cyclical astronomical phenomena linked to the orbits of the Sun, the Moon and the planets \citep{Ptolemy,Masar,Kepler,Swerdlow,Iyengar}. Because of this conviction the ancient astronomers developed calendars that contain several cycles \citep{Aslaksen} as well as the well-known annual cycle.
During the last 70 years, numerous scientific evidences appear to have corroborated that ancient conviction. For example, \cite{Milankovic} theorized  that variations in eccentricity, axial tilt and precession of the orbit of the Earth  determine climate patterns such as  the 100,000 year ice age cycles of the Quaternary glaciation. Milankovitch's theory fits the data very well, over the past million years, in particular   if the temporal rate
of change of global ice volume  is  considered \citep{Roe}.  More recently, a number of authors \citep{Shaviv2003,ShavivV2003,Svensmark2007} have  shown that  the cosmic-ray flux records well correlate with the warm and ice periods of the Phanerozoic during the last 600 million years: in this case  the cosmic-ray flux oscillations are believed to be due to the changing galactic environment of the solar system, as it crosses the spiral arms of the Milky Way.   Over  millennial and secular time scales several authors have found that changes in sunspot number and cosmogenic isotope productions well correlate with climate changes  \citep{Eddy,Sonett,Hoyt,White,vanLoon,Bond,Kerr,Douglass,Kirkby,Scafetta2005,Scafetta2007,Scafetta2008,Shaviv2008,Raspopov,Eichler,Soon,Meehl,Scafetta2009}. Moreover, instrumental global surface records since 1850 appear to be  characterized by a set of frequencies that can be associated to the Moon (9.1-year period) and to the motion of the Sun relative to the barycenter of the solar system (about 10.5, 20, 30 and 60 year periods)\citep{Scafetta2010a,Scafetta2010b}.

In this paper, we study the historical mid-latitude aurora records since 1700 \citep{Krivsky,Silverman} and show that these records share the same set of frequencies that characterize the climate system as well as the natural oscillations of the solar system. This finding reveals the existence of a clearer physical mechanism,  missing in our  previous study \citep{Scafetta2010b}, that could link the astronomical cycles to climate oscillations. The major implication  of this paper is that  there exists  an astronomical harmonic  modulation of the electric properties of the Earth's atmosphere that  modulates cloud cover and, therefore, the terrestrial albedo \citep{Svensmark98,Svensmark2007,Svensmark2009,Kirkby,Tinsley}.

This research would also support the development of a novel astronomically-based theory of climate change that may credibly compete and likely substitute  the current mainstream anthropogenic global warming theory (AGWT)  advocated by the \cite{IPCC}. In fact, the AGWT advocates  claim  that  astronomical forcings of the climate are almost negligible and that the climate variations are induced by some still poorly understood and modeled \emph{internal chaotic dynamics} of the climate system, and by trends in greenhouse gases (GHG) (mostly $CO_2$ and $CH_4$) and aerosol records. More precisely, global surface temperature has risen \citep{Brohan} by about 0.8 $^oC$ and 0.5 $^oC$ since 1900 and 1970, respectively. The \cite{IPCC} and other researchers \citep{Lean} have claimed that more than 90\% of the observed warming since 1900 and practically 100\% of the observed warming since 1970 have had  an anthropogenic cause. While Lean and Rind's methodology  that the climate responds linearly with the forcings can be easily questioned by noting that the heat capacity of the Earth is not zero \citep{Scafetta2009}, the IPCC claims derive from figures 9.5 and 9.6 in the IPCC report (AR4-WG1, 2007)  showing, by means of professional climate general circulation models (GCMs),   that natural forcings alone (volcano and solar irradiance) would have caused a cooling since 1970: thus, the observed post-1970 warming has been interpreted as being induced by human activity alone.
 However, the very large uncertainty in the aerosol forcings and of the climate sensitivity to GHG changes \citep{IPCC,Knutti,Rockstrom,Lindzen,Spencer}, and the current very poor modeling of the water vapor feedback \citep{Solomon2010}, of the cloud system \citep{Lauer}, of the ocean dynamics and of the biosphere  question the robustness  of the current GCMs  for properly interpreting and reconstructing the real climate\citep{Idso}.

 On the contrary, if the climate system is mostly characterized by a specific set of harmonics, it may be possible to partially reconstruct and forecast it in the same way in which ocean tides are currently predicted, that is, by means of harmonic constituent models based on astronomical cycles \citep{Thomson,Fischer,Doodson}.  A harmonic constituent model is just a superposition of several harmonic terms of the type

\begin{equation}\label{eqa1}
    F(t)=A_0 + \sum_{i=1}^N A_i\cos(\omega_i t + \phi_i).
\end{equation}
  The frequencies $\omega_i$ are deduced from astronomical theories. The amplitude $A_i$ and the phase $\phi_i$ of each  harmonic  are empirically determined using regression on an adequate sample of   observations, and then the model is used to forecast future scenarios. Currently, in most US costal locations tidal forecast is made with 35-40 harmonic constituents \citep{Ehret}.

In the following, we  show that  typical energy balance models and general circulation models can be mathematically  reduced to harmonic models in first approximation, and propose that the climate oscillations too can be approximately reconstructed and forecasted by using a planetary harmonic constituent model philosophically equivalent to Eq. \ref{eqa1} based on astronomical cycles.

\section {A possible link between mid-latitude auroras and the cloud system}

 In this paper, we postulate that the annual frequency occurrence of mid-latitude aurora events is a measure of the level of electrification of the global ionosphere, which is mostly regulated by incoming cosmic ray flux variations \citep{Kirkby,Svensmark2007}. When the ionosphere is highly ionized by cosmic rays, large auroras would more likely form at the mid-latitudes. This phenomenon would occur because when the upper atmosphere is highly ionized, it would also be electrically quite sensitive to large solar wind particle fluxes and favor the formation of extended mid-latitude auroras. In fact, higher ionization of the atmosphere would mostly occur when the magnetosphere is weaker and cosmic ray as well as solar wind particles, can more easily reach the mid-latitudes. Then, the level of atmospheric ionization and of the global electric circuit of the atmosphere should regulate the cloud system \citep{Kirkby,Svensmark2007,Tinsley}. If the above theory is correct, the frequencies of the mid-latitude aurora records should be present in the climate records too.

Indeed, cloud-related climatic effects can largely dominate other mechanisms such as $CO_2$ and $CH_4$ GHG forcing \citep{Kirkby,Svensmark2007}. For example, in the past billion years the Earth experienced severe glaciations despite the fact that the $CO_2$ concentration was at least 10 times higher than today (4000-6000 ppmv against the actual 390 ppmv) \citep{Hayden}. In addition to major continental drifts and other  geological  events \citep{Courtillot}, it has also been argued that the high cosmic ray incoming flux that occurred during known major glaciations  could have increased the cloudiness of the Earth causing a global cooling \citep{Shaviv2003,Shaviv2008,ShavivV2003,Svensmark2007}.

  The cloud system controls a large part of the terrestrial albedo and regulates the amount of total solar irradiance reaching the Earth's surface. The solar irradiance reaching the Earth's surface   warms the ocean and the land. A small astronomical modulation of the terrestrial albedo through the cloud system can greatly increase the climate sensitivity to solar forcing \citep{Shaviv2008,Scafetta2009}.  Evidently, if the current GCMs are missing an important forcing of the cloud system, they would poorly reconstruct and, potentially, severely misinterpret climate variations at multiple temporal scales.  Indeed, a correlation between galactic cosmic ray fluxes and high-altitude, mid-altitude and low clouds have been found \citep{Svensmark2007,Rohs,Laken}.

\begin{figure}[t!]
\includegraphics[angle=-90,width=21pc]{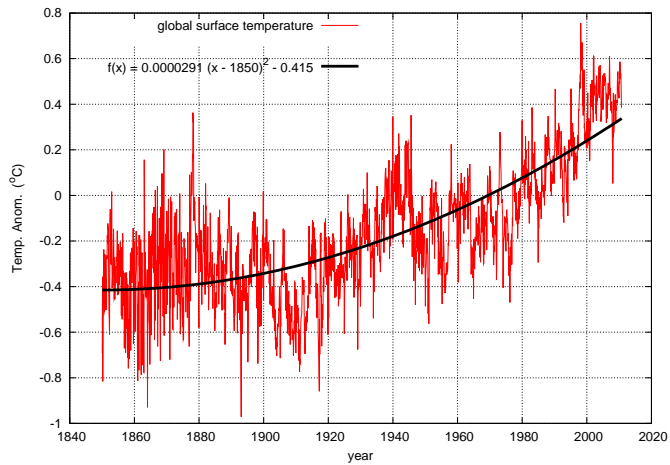}
\caption{ Global surface temperature anomaly  \citep{Brohan}. The figure also shows the quadratic fit upward trend of the temperature (black).  }
\end{figure}

\section{A 60-year cycle in mid-latitude aurora and climate records}

Figure 1 shows the global surface temperature (HadCRUT3) \citep{Brohan} from 1850 to 2010 (monthly sampled). Global surface temperature records (land, ocean, north hemisphere, south hemisphere) have been found to be characterized by a clear and large quasi 60-year cyclical modulation \citep{Scafetta2010b}. In fact, the following 30-year trends are evident in the record: 1850-1880, warming; 1880-1910, cooling; 1910-1940, warming; 1940-1970, cooling; 1970-2000, warming; and a small cooling since 2000 that may last until 2030-2040.

\begin{figure}[t!]
\includegraphics[angle=0,width=21pc]{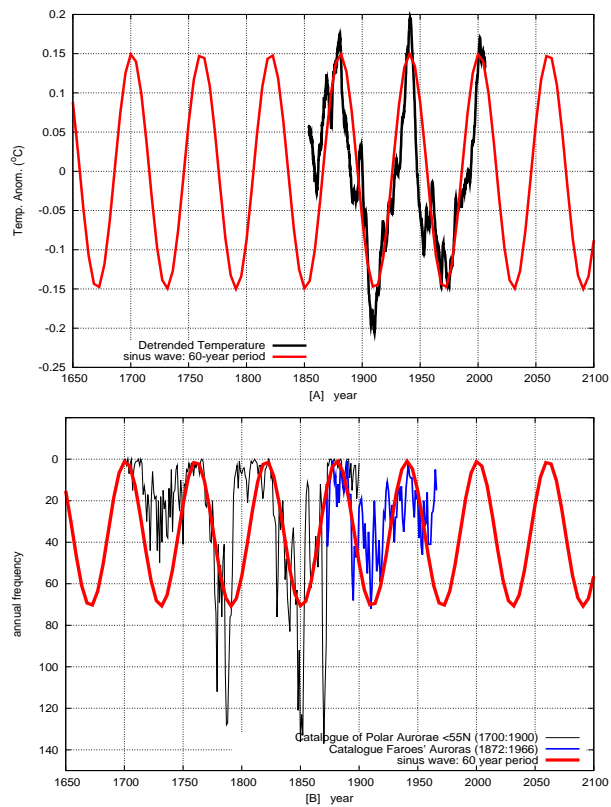}
\caption{ [A] The 60-year cyclical modulation of the global surface temperature obtained by detrending this record of its upward trend shown in Figure 1. The temperature record has been filtered with a 8-year moving average. Note that detrending a linear or parabolic trend does not significantly deform a 60-year wave on a 160-year record, which contains about 2.5 of these cycles, because first and second order polynomials are sufficiently orthogonal to a record containing at least two full cycles. On the contrary, detrending higher order polynomials would deform a 60-year modulation on a 160-year record and would be inappropriate. [B] Aurora records from the ``Catalogue of Polar Aurora $<$55N in the Period 1000-1900'' from 1700 to 1900 \citep{Krivsky}.  Figure 2B also depicts the catalog referring to the aurora observations from the Faroes Islands from 1872 to 1966. Both temperature and aurora records show a synchronized 60-year cyclical modulation as proven by the fact that the 60-year periodic harmonic function superimposed to both records is the same.}
\end{figure}

Figure 2A shows the global surface temperature record detrended of its upward trend and smoothed with a 8-year moving average that highlights its  60-year cyclical modulation with a peak-to-trough amplitude of about 0.3-0.4 $^oC$.

Figure 2B shows the annual frequencies of mid-latitude auroras obtained from the supplement of the catalogue of mid-latitude auroras $<$55N from 1700 to 1900. This record contains the historical aurora observations reported mostly in central Europe since 1000 AD \citep{Krivsky}.  Before 1700, the record is largely incomplete and the data are not depicted in the figure. Figure 2B also depicts the catalog of the aurora observations in the Faroes Islands from 1872 to 1966. Despite the fact that the Faroes' record refers to a northern region (62N), \cite{Silverman} noted that it appears to have physical properties compatible with the Krivsky and Pejml's record. So, it may be reasonable to combine the two catalogs covering 266 years from 1700 to 1966. The present author does not know why these two records stop in 1900 and 1966 respectively. Aurora events surely have be collected but not with the same naked eye methodologies used in the past. The most recent electronically based records could show technical heterogeneities with the historical ones such that the old and recent records cannot be simply combined. The aurora records are flipped up-down and superimposed to the same 60-year periodic sinusoidal function depicting the 60-year temperature cycles in Figure 2A.  Mid-latitude auroras are more likely to occur during cooler multidecadal periods.

An approximate 60-year modulation in the aurora record from 1000 to 1900 has been observed by \cite{Charvatova1988}, once that the original record before 1700 is adjusted with what is there called a ``civilization'' factor coefficient. \cite{Komitov} also found a correspondent quasi 60-year cycle in both aurora and Greenland and Antarctic $^{10}Be$ concentration records during the period 1700-1900.
Periodic patterns and long-term variations in the aurora records have been observed for centuries since the work by \cite{Mairan}. De Mairan found several reprises and returns in the aurora activity. \cite{Siscoe} summarized a number of the studies that searched for periodicities in aurora occurrence. For example, in 1831 Hansteen inferred a period of 95 years; \cite{Olmsted}
found distinctive evidences for 60 to 65 years as the intervals between great auroras; and Wolf and Fritz preferred a period of about 55 years. Another work also indicated periods in the region of 80-90 years \citep{Feynman}, and even large periods
as long as 200 or 400 years.  Large cycles with periods of about 60, 80-90 years and longer bi- and multi-secular cycles are the most commonly reported. These frequencies correspond to the well-known Gleissberg and Suess solar cycle frequency bands \citep{Suess,Krivsky,Ogurtsov}.

\begin{figure}[t!]
\includegraphics[angle=0,width=21pc]{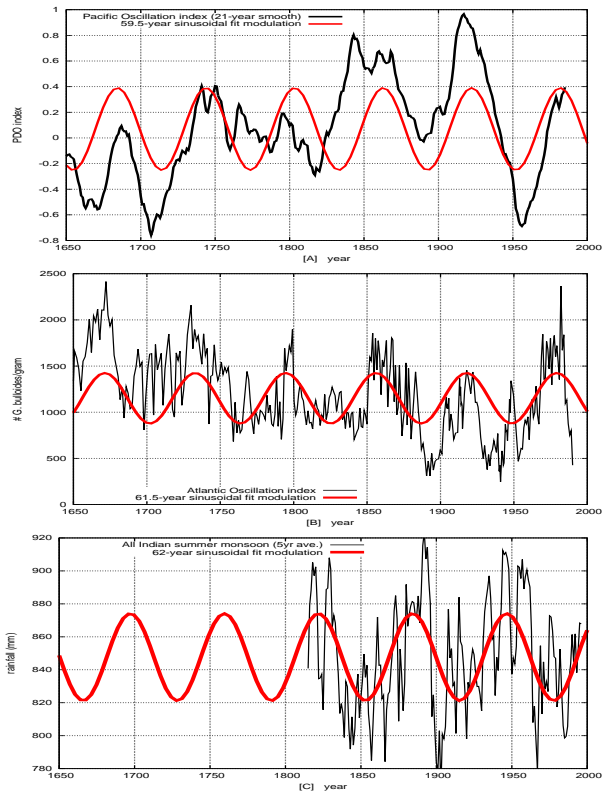}
\caption{
[A] Twenty-year moving average of the tree-ring chronologies from Pinus Flexilis in California and Albertain: this record is used as a proxy for reconstructing the Pacific Decadal Oscillation \citep{MacDonald}.
[B] Record of G. Bulloides abundance variations (1-mm intervals)   from 1650 to 1990 A.D. (black line) \citep{Black}; this is a  proxy for the Atlantic variability since 1650.
[C] Five-year running average of the Indian summer
monsoon rainfall from 1813 to 1998 \citep{Agnihotri2003}. All three records show clear 60-year cyclical modulations that are (positively or negatively) well correlated to the 60-year cycles of the global surface temperature and the aurora records. The records are best fit with sinusoidal functions that give a statistical error about the 60-year period of $\pm 4$ years.  }
\end{figure}

As already discussed in \cite{Scafetta2010a,Scafetta2010b}, quasi 60-year cycles are found in several climatic from several regions of the Earth \citep{Klyashtorin2001,Klyashtorin2007,Klyashtorin2009,Courtillot44,Camuffo,Agnihotri2002,Agnihotri2003,Sinha,Goswami,Yadava,Mazzarella1,Jevrejeva} and astronomical/solar records \citep{Yu,Patterson,Ogurtsov,Roberts}.
 These results clearly suggest the existence in the climate of a natural 60-year cycle synchronized to a correspondent   solar/astronomical cycle. For example, Figure 3 depicts three multi-secular climate records from three independent regions of the globe showing multiple quasi-60 year large oscillations since 1650.  Figure 3A depicts a record obtained from the tree-ring chronologies from Pinus Flexilis in California and Albertain: the best sinusoidal fit gives a period of $T=61.5 \pm 4$ years. This record is used as a proxy for reconstructing the Pacific Decadal Oscillation \citep{MacDonald}.
 Figure 3B depicts the G. Bulloides abundance variation record found in the Cariaco Basis sediments in the Caribbean sea since 1650 \citep{Black}: the best sinusoidal regression fit gives a period of $T=59.5\pm 4$ years: a quasi-60 year cycle has been found in the Caribbean sea for millennia during the entire Holocene  \citep{Knudsen}. This record is an indicator of the trade wind strength in the tropical Atlantic Ocean and of the North Atlantic Ocean atmosphere variability. Periods of high G. Bulloides abundance correlate well with periods of reduced solar output (the well-known Oort, Wolf, Sp\"orer, Maunder, Dalton minima): a fact that suggests that these cycles are solar induced \citep{Black}.
 Figure 3C depicts the Indian summer monsoon rainfall record from 1813 to 1998 years that also shows prominent quasi 60-year cycles. The Indian summer monsoon rainfall record, together with east equatorial and Chinese monsoons,  clearly manifest a solar variability signature for several centuries \citep{Agnihotri2003}:  the best sinusoidal fit of the Indian monsoon rainfall record gives a period of $T=62\pm4$ years.  The cycles depicted in Figures 2 and 3 are well synchronized to the quasi 60-year modulation of the global temperature observed since 1850.

Consequently,  the very good correspondence between the two 60-year periods 1880-1940 and 1940-2000, which are observed in all (North and South, Ocean and Land) global surface temperature records \citep{Scafetta2010a,Scafetta2010b}, is unlikely just a coincidental red-noise pattern. With a high confidence level, a natural 60-year periodic modulation exists in the climate system because these cycles are observed in a large number of natural multi-secular records although this cycle may not appear always evident because of errors in the data and because of other superimposed patterns of different physical origin (for example, volcano effects) and other natural cycles that could induce interference patterns.

\section{Spectral analysis of aurora and climate data, and joint power statistical tests}

To determine whether faster cycles are equally common in the aurora and temperature records we analyze the global surface temperature and the auroras records by using the maximum entropy method \citep{Ghil} because with a proper high number of \emph{poles} this technique resolves the very low frequency band of the spectrum much better than Fourier or periodogram based techniques  \citep{Priestly}. Note that in our case  MEM needs to be used with a very large number of poles (up to half of the length N of the record) because we are interested in resolving the very low frequency range of the spectrum where the dacadal and multidecadal periodicities  are located, as simple computer experiments with synthetic harmonic records would easily prove.
A number of poles between N/5 and N/2 (at most)  is correct in most cases \citep{Ulrych,Courtillot1977}.

\begin{figure}[t!]
\includegraphics[angle=0,width=21pc]{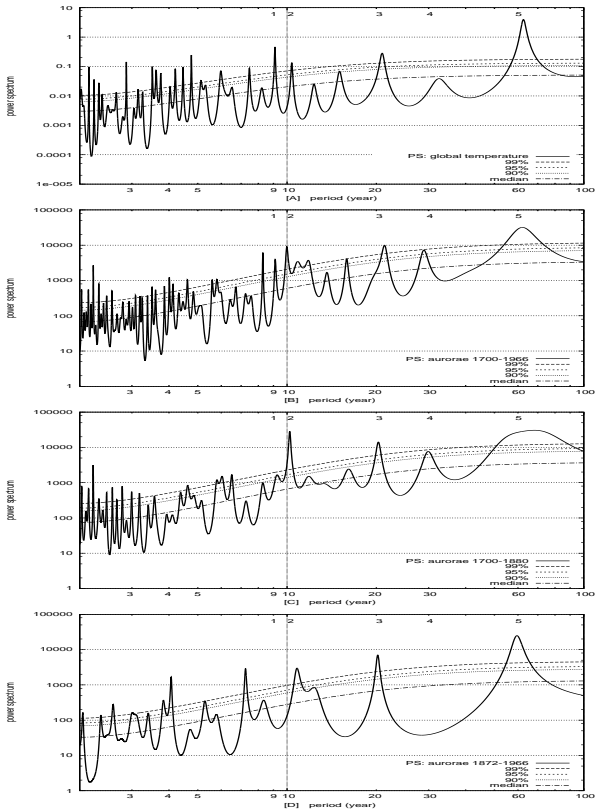}
\caption{ [A] Power spectrum of the global surface temperature [A], which covers the period 1850-2010. [B,C,D] power spectra of three auroras records, which cover the periods 1700-1966, 1700-1880 and 1872-1966 respectively. The numbers \#1, \#2, \#3, \#4 and \#5 indicate the periods at about 9, 10-11, 20, 30 and 60 years. These peaks are statistically significant against red-noise background tests at different confidence levels (median, \% 90, \%95, \%99), which are indicated with the dot lines.   }
\end{figure}

Figure 4A shows the power spectrum of the temperature that covers the period 1850-2009. This curve corresponds to those analyzed in \cite{Scafetta2010b}, where additional alternative statistical tests to check the robustness of the results are conducted. Four major decadal and multidecadal cycles are seen at about 9, 10-11, 20-21 and 60-year periods. These cycles are indicated with the number \#1, \#2, \#3 and \#5. With the exception of the smaller peak \#4 (about 30 years), which may not be significant, all other four frequencies have a 99\% confidence against red noise background. Faster cycles or less important cycles are not studied here. Traditional periodogram analysis (not reported in this figure) produces the same major four peaks detected by MEM, but with a larger uncertainty, and confirms that the global surface temperature is characterized by at least the above four major frequency peaks.

Figure 4B shows the power spectrum of the superposition of the catalogues of mid-latitude auroras merged with the Faroes' auroras: this combined record covers the period 1700-1966. Again, five peaks cycles are clearly seen at about 9, 10-11, 20-21, 30 and 60 years.
Figure 4C shows the power spectrum of a portion of the catalogue of mid-latitude aurora covering the period 1700-1880, which presents a minimum overlap with the temperature data and with the Faroes' catalog. Again, five peaks are seen at about 9, 10-11, 20-21, 30 and 60 years.
Figure 4D shows the power spectrum of the Faroes' catalog that covers the period 1872-1966. In this case, only 3 cycles are clearly seen: cycle \#2 (10-11 years); cycle \#3 (20-21 years); and cycle \#5  (about 60 years).
Note that cycle \#1 (about 9 years) is not as strong as the one found in the temperature record: we will discuss this finding later.

Table 1 reports the measured five periods and their average values. The frequencies of the temperature and of the AAR columns are compatible within their error of measure. For the 5 period couples reported in Table 1, the reduced $\chi^2$ is $\chi^2_o=0.18$ with 5 degrees of freedom and the coherence probability is $P_5 (\chi^2 \geq \chi^2_o)>96\%$.

\begin{figure}[t!]
\includegraphics[angle=-90,width=21pc]{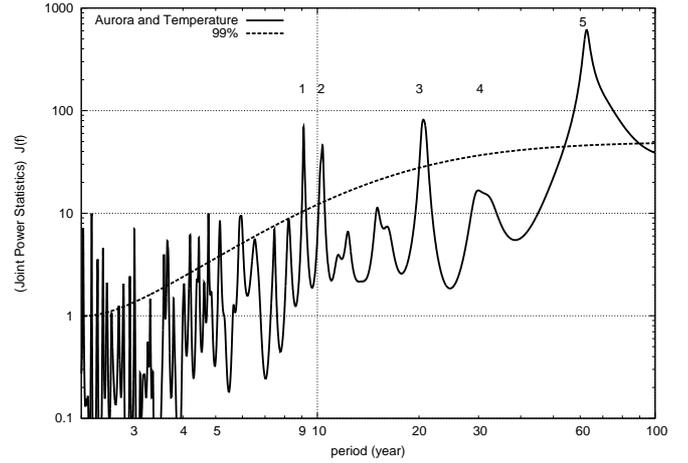}
\caption{ Joint power statistics \citep{Sturrock} between the power spectra of the 1850-2010 global temperature (Figure 4A) and the 1700-1900 aurora record (Figure 4C). The joint peaks at periods of about 9, 10.5, 20 and 60 years are evident and have a 99\% confidence level against red-noise background.}
\end{figure}

Figure 5 further supports the result that the aurora records and the global temperature record share a similar set of frequency. In this case, we use the joint power statistic function \citep{Sturrock} which confirms that the aurora record and the global temperature record share a common set of major cycles  at about 9, 10.5, 20, and 60-year periods (cycles \#1, \#2, \#3 and \#5) with a 99\% confidence. If two power spectra share a compatible frequency peak, also their joint power statistic function, which is approximately given by $\sqrt{S_1(f)*S_2(f)}$, would show a significant peak at the same frequency.  Other common cycles may be present as well, but they are less evident.

\section{A possible astronomical planetary origin of the aurora and temperature oscillations}

The observed aurora and climate cycles suggests that the solar system planetary orbital dynamics may be their first physical cause. Indeed, some studies \citep{Wolf1859,Schuster,Bendandi,Jose,Fairbridge,Landscheidt1988,Landscheidt1999,Charvatova1990,Charvatova2000,Charvatova2009,Charvatova2004,
Grandpierre,Mackey,Hung,Wilson2008,Sidorenkov,Perryman}   have suggested that solar variation can be partially driven by the  planets through  gravitational  spin-orbit coupling mechanisms and gravitational tides.

A full theory that would physically explain how the solar wobbling or the planetary tides may influence solar activity has not been fully developed yet. However, preliminary studies suggest that planetary gravity may increase nuclear rate \citep{Grandpierre,Wolff} by favoring the movement of fresh fuel into the solar core. The proposed mechanisms would likely produce the major frequencies herein discussed because it is based on the study of the wobbling of Sun around the solar system barycenter as done in \cite{Scafetta2010b}.

Moreover, solar wobbling could be physically relevant relative to the incoming cosmic ray flux. In fact, because the inner part of the solar system is almost gravitationally locked to the Sun, and the Earth's orbit too wobbles around the solar system's barycenter almost in the same way as the Sun does. Indeed, an observer from the Earth does not see the solar wobbling while an observer external to the solar system would see the Earth's orbit wobbling almost in the same way as the Sun does. Thus, relative to the Sun or to the Earth's orbit, the incoming cosmic ray flux is wobbling with the same frequencies found in the solar barycentric motion \citep{Scafetta2010b}.

The major planetary resonances of the solar system related to Jupiter and Saturn are the following:  Jupiter's sidereal period 11.862 years;  Saturn's sidereal period  29.457 years;  the opposition-synodic period of Jupiter and Saturn (9.93 year), which oscillates between 9.5 and 10.5 years due to the eccentricity of their orbits;   about 19.84 years, the synodic period of Jupiter and Saturn;  about 59.6 years, the repetition of the combined orbits of Jupiter and Saturn; and a spring tidal beat frequency ($\sim$61 year).
Moreover, there is the 11-year Schwabe solar cycle and the 22-year Hale magnetic solar cycle which perhaps are indirectly associated to the two major Jupiter/Saturn tides because the 11-year cycle appears constrained between the 9.3-year spring Jupiter/Saturn tide and the 11.8-year Jupiter tide \cite{Wilson1987}. Thus, three major solar and solar system oscillations should be expected at periods of about 10-11, 20-22 and 59-63 years. Many other resonance frequencies related to all planets have been found in the Sun's motion \citep{Bucha,Charvatova1988}, but they are not discussed here.

An additional large 9.1-year temperature cycle does not appear among the planetary resonances and \cite{Scafetta2010b} found that it may be a solar/lunar tidal cycle. In fact, a 9.1-year periodicity is between the period of the recession of the line of lunar apsides, about 8.85 years, and half of the period of precession of the luni-solar nodes, about 9.3 years (the luni-solar nodal period is 18.6 years). The 9.1-year cycle is also about half of the 18.03-year Saros eclipse cycle. Note that every almost 9 tropical years the Sun, the Earth and the Moon are aligned at the same angle relative to the Earth's surface. Because two opposite tides are simultaneously present, the two alignments Sun-Moon-Earth and Sun-Earth-Moon are physically equivalent.  Consequently, a quasi 18-year astronomical solar/lunar cycle is associated to a quasi 9-year tidal cycle dynamics.

The results depicted in Figures 4B, 4C and 4D suggest that the aurora events are synchronized to the above astronomical periodicities.
About the two decadal cycle in the climate records, several studies have attempted to interpret it either as a lunar influence on climate via the 18.6-year luni-solar nodal or the nutation cycle, or as an influence of the 22-year Hale solar magnetic cycle \citep{Currie,Keeling,Camuffo2001,McKinnell,Munk,Hoyt}. However, on a global scale the most prominent bi-decadal cycle has a period of about 20-21 years as shown above in Figures 4 and 5, while a distinct  18.6 year lunar cycle or a 22-year solar cycle are not clearly visible in the adopted global temperature records. Perhaps, a 18.6-year nutation cycle may become particularly significant only in some regions, such as the polar ones. Because  the global data herein analyzed appear to show a quasi 20-21 year cycle more than a sharp 18.6 or a 22-year cycle, we claim that a quasi 20-21 year cycle is what may characterizes the well known bi-decadal climatic cycle \citep{Scafetta2010b}.

\begin{figure}[t!]
\includegraphics[angle=0,width=21pc]{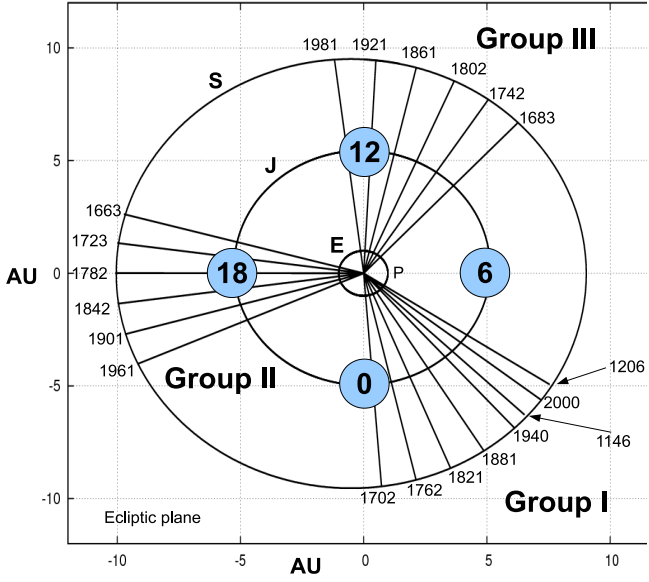}
\caption{ Dates and positions of the conjunctions between Jupiter and Saturn from 1650 to 2010, as reported in Table 2. The figure depicts the orbits of the Earth (E), of Jupiter (J) and of Saturn (S). `P' is the position of the perihelion of the Earth. The figure emphasizes the tri-synodic cycle of Jupiter and Saturn. It takes about 60 years for Jupiter and Saturn to reach the same relative alignment around the sun. Two J/S conjunctions occurred in 1146 and 1206 are added. The latter two J/S conjunctions reveal a 800-850 year cycle in the J/S conjunction circulation as seen from the Earth and noted by Kepler who was using the tropical orbital period of the planets.    }
\end{figure}

\section{Further analysis of Jupiter and Saturn's orbital dynamics and of their induced solar system cycles}

Herein we further analyze how the orbital eccentricities and the periods of Jupiter and Saturn can produce major 10, 20 and 60-year oscillations in the solar system. Figure 6 shows the dates of all conjunctions of Jupiter and Saturn occurred from 1650 to 2010 relative to the Sun. The orbital data are obtained using the NASA Jet Propulsion Laboratory Developmental Ephemeris. The angular separation between Jupiter and Saturn relative to the Sun at the conjunction day is of the order of 1$^o$, or below. Table 2 reports the dates, the heliocentric coordinates and the distance from the Sun of the conjunctions of Jupiter and Saturn. The coordinates are chosen to be the right ascension and declination, which refer to the equatorial plane of the Earth. Thus, the declination measures the direction of the gravitational and magnetic forces of Jupiter and Saturn relative to the daily average orientation of the terrestrial magnetosphere.
The figure shows that the conjunctions since 1650 can be separated into three groups. A conjunction in the Group I is followed by a conjunction in the Group II at an angle of about $360^o*19.86/29.46 =242.7^o$ after about 20 years, which is followed by a conjunction in the Group III again at about 242.7$^o$, after about 20 years. Every 3rd alignment (about 60 years) is about 8.1 degrees of the original starting point: the configuration repeats about every 900 years. A similar diagram was prepared by \cite{Kepler} to  interpret the  Star of Betlehem as a J/S conjunction \citep{Kemp}, but Kepler used the tropical orbital periods that yield an angular separation of about 3 degree: and the configuration would repeat approximately every 800 years, as depicted in Figure 6.

 The last column of Table 2 reports the approximate magnitude in Km of the tidal elongation produced by the combined effect of Jupiter and Saturn at 1 AU from the Sun, which is the Sun-Earth average distance. The tabulated spring tidal elongation values at 1 AU from the Sun are calculated by using the following approximated formula \citep{Taylor}:

\begin{equation}\label{eqte}
    a_{+} - a_{-}=\frac{3}{2}\frac{M_J}{M_{Su}}\frac{R_{SuE}^4}{R_{SuJ}^3} +
    \frac{3}{2}\frac{M_S}{M_{Su}}\frac{R_{SuE}^4}{R_{SuS}^3},
\end{equation}
where: $a_{+}$ and $a_{-}$ are, respectively, the maximum and minimum radius of the equipotential surface spheroid produced by the gravitational field of the Sun plus the tidal potentials of Jupiter and Saturn; $M_{Su}=333000$, $M_J=317.9$ and $M_S=95.18$ are the masses of the Sun, Jupiter and Saturn in Earth's masses;  $R_{SuJ}$ and $R_{SuS}$ are the distances of the Sun from Jupiter and Saturn, respectively; and $R_{SuE}=1$ AU is the average Sun-Earth distance.

Table 2 shows that the tidal elongations at the J/S conjunctions occurring in Group I are significantly larger (about 1800 Km) than those occurring in the other two groups: about 1400 Km and 1600 Km respectively.  The results about the tidal elongation indicate that Group I alignments occur when Jupiter and Saturn occupy the closest positions to the Sun and their combined tidal effect is the largest.  This returning pattern gives origin to a 60-year physical tidal cycle on the Sun and in the heliosphere in proximity of the Earth's orbit. A similar 60-year periodic pattern would be generated by the magnetic fields of Jupiter and Saturn that influence the heliosphere too.

To show the varying effect of the tidal combination of the J/S synodic cycle and Jupiter eccentricity cycle we observe that: a) the tidal oscillation produced by Jupiter and Saturn orbits at 1 AU have an amplitude of about 224 km and 12.3 Km, respectively, which are calculated as half of the difference between the tidal elongation at the perihelion and at the aphelion; b) J/S synodic cycle produces an average tide amplitude of about 38.4 Km, which is the half  amplitude of the average Saturn induced tidal elongation. Jupiter's perihelion occurred on 20/May/1999 ($\tau_1=1999.38$), Saturn's perihelion occurred on 27/Jul/2003 ($\tau_2=2003.57$) and the J/S conjunction occurred on 23/Jun/2000 ($\tau_3=2000.48$). Thus, the average J/S tidal anomaly effect approximately oscillates as

\begin{eqnarray}
\label{Tt}
 D(t) &=& 224 \cos \left[\frac{2\pi~(t-\tau_1)}{11.86}\right] +  \\ \nonumber
   & & \left( 38.4 + 12.3\cos \left[ \frac{2\pi~(t-\tau_2)}{29.46} \right] \right) ~\cos \left[ \frac{2\pi~(t-\tau_3)}{9.93} \right],
\end{eqnarray}
where it is assumed the superposition of two waves given by the tidal sidereal orbit of Jupiter and the spring tidal oscillation of Jupiter and Saturn.

\begin{figure}[t!]
\includegraphics[angle=0,width=21pc]{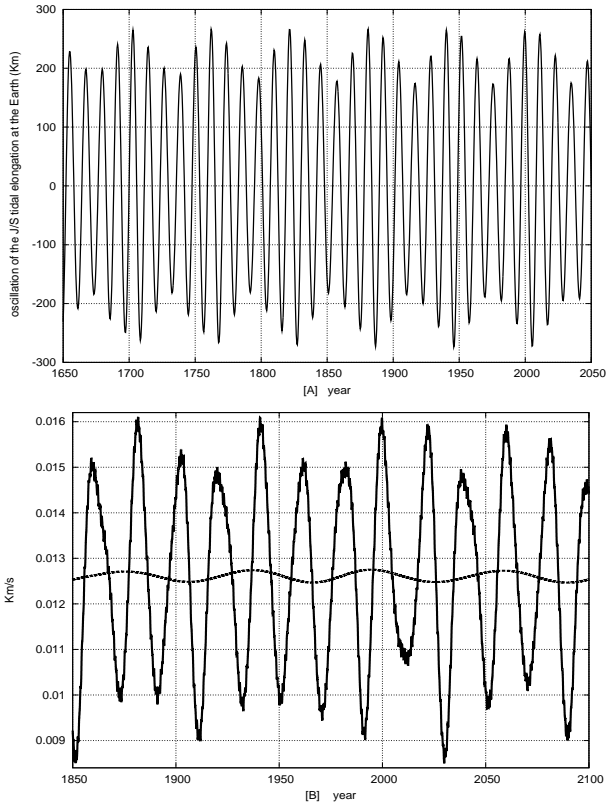}
\caption{[A] The Jupiter-Saturn tidal anomaly as deduced from Eq. \ref{Tt} at 1 AU from the Sun. Note the 60-year beat modulation. This 60-year modulation is in good phase with the 61-year cycle of the temperature and of the aurora records since 1700 depicted in Figures 2 and 3: the largest magnitudes correspond to peaks of the 61-year cycles. [B] Figure B depicts the speed of Sun relative to the solar system barycenter (SSSB) (Scafetta, 2010b). The 20-year (solid) and 60-year (dash) oscillation is evident. The dash curve is a moving average filtering. Note that the 60-year modulation observed in both figures corresponds exactly to the 60-year modulation of the auroras and of the climate depicted in Figure 2. }
\end{figure}

Figure 7A depicts the function $D(t)$ for the period 1650-2050. The quasi 60-year beat modulation is evident, and it is in good phase with the 60-year modulations observed in Figures 2 and 3. Note, for example, that the function $D(t)$ presents minimum amplitudes during periods of temperature minima (1850-1860, 1900-1920, 1960-1980) and maximum amplitudes during periods of temperature maxima (1870-1890, 1930-1950, 1990-2010).

Figure 7B depicts the speed of Sun relative to the barycenter of the solar system where the 20-year and 60-year modulation of the solar orbit is also evident. Note that the 60-year modulation of the solar speed corresponds exactly to the 60-year modulation of the auroras and of the climate depicted in Figure 2.

\begin{figure}[t!]
\includegraphics[angle=-90,width=21pc]{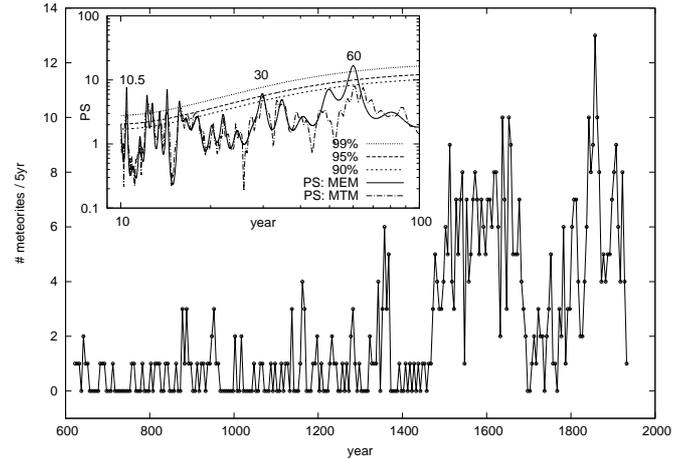}
\caption{Number of historically recorded meteorite fall in China from AD 619 to 1943 \cite{Yu}. Note that the gradual multi-secular increase of witnessed meteorite falls is likely due to social parameters effects. In fact, in China, the typography (book-printing) has been known since AD. 1041 while only hand written records are available before, so the records are very scarce in the first millennium in comparison with the greater number of records in the second half of the second millennium. However, this trending should not significantly alter the faster decadal and multidecadal modulation, which should be physical meaningful. The insert shows the power spectrum of the record with the Maximum Entropy Method (50 poles) and the Multi Taper Method. Cycles at 10.5, 30 and 60 years are clearly visible.}
\end{figure}

The existence of an astronomical 60-year cycle that is influencing the inner part of the solar system is further proven by meteorite fall records, as extensively studied in \cite{Yu}. It is well known that Jupiter and Saturn can regulate the collision rate of small bodies with the Earth \citep{Bennett}. Figure 8 shows our analysis of a number of historically recorded meteorites falls in China from AD 619 to 1943. Although this record may be incomplete and should be corrected by some social parameters, it shows cycles at about 10.5, 30 and 60 year periods.  This result would stress the importance of the J/S 60-year cycle in driving a major astronomical oscillation of the solar system.  The power spectrum in the insert of Figure 8 also suggests the astronomical origin of a significant $\sim$10.5-year cycle, which is also found in the temperature record (see Table 1) and in the alignments of the tide-raising planets within ten degrees \citep{Okal,Bucha,Grandpierre}.

Quasi-millennial solar cycles are evident in several cosmogenic isotope productions \citep{Bard,Bond,Kerr,Steinhilber}.
 Figure 6 shows that the J/S conjunctions in AD 1146 and 1206 occurred approximately at the same position of the conjunction in 2000. Thus, the J/S conjunction line presents a 800-860 year cycle, as \cite{Kepler} found where the tropical orbital periods ($P_J=11.857$ and $P_S=29.424$) were used. Moreover, the distance of the two planets from the Sun at the conjunction dates presents a quasi 900-year cycle because of their orbital precession, as it is evident by comparing the tidal elongations in 1086 and 2000, which could be obtained using the sidereal orbital periods ($P_J=11.862$ and $P_S=29.457$), which are more physical relative to the Sun. The orbital pattern reveals the existence in the solar system of quasi-millennial cycles. Quasi-millennial cycles have been found in solar proxy records [for example see: \cite{Hood,Charvatova2000,Bond,Kerr,Steinhilber}] and quasi 900-year cycles have been found in the Holocene climate variability \citep{Schulz}. A quasi-millennial cycle is quite evident in numerous global temperature reconstructions that show a large preindustrial variability. These reconstructions show a distinctive Roman Optimum (0-400 AD), a Medieval Dark Age (400-900 AD), Medieval Warm Period (MWP) (900-1300 AD), a Little Ice Age (LIA) (1400-1800 AD) and a Current Warm Period (CWP) since 1800 \citep{Moberg,Loehle,McShane,Ljungqvist2010} and in numerous recent regional records  such as from England, Sargasso sea, Japan, Antarctic, Canada, Chile, Indo-pacific region, China, tropical South America, and many more \citep{Lamb,Parker,Keigwin,Adhikari,Khim,Huang,Cook,vonGunten,Oppo,Ge,Kellerhals,Mann2008,Ljungqvist2009}. However, these quasi-millennial cycles, as well as other secular and multi-secular cycles  may be the topic of another work; we only note that the latter temperature reconstructions contradict the \emph{Hockey Stick} temperature graphs \citep{Mann2003} that do not show a significant MWP nor a LIA, but only an unprecedented CWP. The most recent temperature reconstructions would imply  the overall solar forcing of the  climate  has been  underestimated by at least a factor of three by traditional energy balance models \cite{Scafetta2010a}.

\section{Cloud cover oscillation as a possible mechanism for climate cycles}

  Total solar irradiance variations alone are usually claimed to be too small to induce large climate variations \citep{IPCC}. Additional solar related forcings would be necessary to explain climatic variations.  Indeed, there may be  an additional effect on climate due to ultraviolet modulation of the stratospheric ozone and stratospheric water vapor \citep{Stuber,Solomon2010}, and a modulation of the cloud cover due to cosmic ray flux variation and  a variation of the global electric circuit \citep{ShavivV2003,Svensmark2007,Rohs,Laken,Tinsley}.

 Variations of the frequency of mid-latitude aurora reasonably reveal electric variations occurring in the terrestrial magnetosphere and ionosphere.  Likely, when the upper atmosphere is more ionized, it is more sensitive to solar wind peak activity and mid-latitude aurora events occur more frequently.   Therefore, the mid-latitude aurora records can be considered as a proxy of the variation of the atmospheric ionization and, likely, of the atmospheric global electric circuit. As proposed by several authors \citep{Tinsley,Svensmark2007}, changes in the global electric circuit can drive changes of cloud formation and cover. A change of cloud cover alters the albedo and can potentially cause a significant climate change. Because auroras oscillated with planetary frequencies the climate would oscillate with the same frequencies  and would be almost perfectly synchronized to the 10-11,  20-21 and 60-62 year cycles of the J/S conjunction, tidal and solar cycles, as shown in \cite{Scafetta2010b}. Several other cycles are surely present, but we do not discuss them here.

To emphasize the effect that a small variation of the albedo may have on the climate sensitivity to solar changes we can do the following exercise. Let us write:

\begin{equation}\label{yu67}
\Delta T \approx \sum_F \Delta T_F \approx \sum_F k_F ~\Delta F,
\end{equation}
where $\Delta T_F$ is the change in temperature caused by a small change of a generic  forcing $\Delta F$.
The value $k_F$ is the equilibrium climate sensitivity to a small changes in the forcing, as measured by $\Delta F$. In the case of solar irradiance the above equation is substituted with $\Delta T_S \approx k_S ~\Delta F_S$. The same can be done for any other  forcings.

By differentiating directly a corrected Stefan-Boltzmann's black-body equation

 \begin{equation}\label{rtrtrt}
f~\frac{(1-a)I}{4}=\sigma T^4,
\end{equation}
 we get

 \begin{equation}\label{eqbs}
    k_S=\frac{\Delta T}{\Delta I}=\frac{T}{4I}=0.053 ~K/Wm^{-2}
 \end{equation}
  using $I=1360$ $W/m^2$ as the average total solar irradiance, $a=0.3$ as the average albedo, $T=289$ K as the global average temperature of the Earth's surface  and  $\sigma=5.67\times10^{-8}$ $W/m^2K^4$ is the Stefan-Boltzmann's constant. A corrective amplification coefficient $f=1.65$   is used to approximate the greenhouse effect of the terrestrial atmosphere to let Eq. \ref{rtrtrt} to give an equilibrium temperature of $T=289$ K.

The above $k_S$ estimate in Eq. \ref{eqbs} assumes, for example, that the albedo \emph{a} is   constant. Now, let us assume that an increase of the total solar irradiance input of $\Delta I=1$ $W/m^2$ is accompanied at equilibrium with a decrease of the albedo by just 1\% that could be caused also by alternative related solar forcing such as a cosmic ray flux solar modulation.  The decrease of the albedo may be due to a decrease of the low cloud cover and other possible factors such as a darkening of the Earth's surface due to a melting of the glaciers and biosphere changes. If we assume that everything else remains constant, the climate sensitivity to total solar irradiance change becomes

  \begin{equation}\label{121212}
k_S =\frac{\Delta T}{\Delta I}=\frac{\sqrt[4]{\frac{(1-a*0.99)(I+\Delta I)f}{4\sigma}}- \sqrt[4]{\frac{(1-a)If}{4\sigma}}}{\Delta I}  \approx 0.36 ~K/Wm^{-2},
\end{equation}
  which is seven times  larger than the value found in Eq. \ref{eqbs} and would be approximately compatible with the empirical values found in \cite{Scafetta2009}. This suggests that even very small solar-induced modulation of the albedo (just a 1 or 2\% variation) can greatly amplify the climate sensitivity to solar irradiance changes even by one order of magnitude and could easily induce climatic oscillation of the order of a few tenths of Celsius degree, which would be perfectly compatible with the observations.

There may also be the possibility that the terrestrial albedo is directly modulated by other astronomical mechanisms such as a compression of the Earth's magnetosphere due to tidal forces.  Also in this case Eq. (\ref{rtrtrt}) would present an albedo that would be a function of time, $a(t)$, and this function would vary with the periods of the astronomical forcing. Even in the eventuality that the incoming total solar irradiance $I$ and the greenhouse feedback coefficient $f$ remain constant, it is easy to prove, using Eq. (\ref{121212}), that a reduction of the albedo by just 1\%, for example a decrease from $a=0.300$ to $a=0.297$, would cause an increase of the global temperature at equilibrium of about $0.3$ K.

 The oscillations observed in the aurora records may suggest that the cloud cover oscillates in the same way. Thus, we can assume, in first approximation, an albedo function that changes periodically as

   \begin{equation}\label{albedo}
a(t) = a_0  + \sum_i a_i~\cos\left(\omega_i~t+\phi_i\right),
\end{equation}
 where $a_0 \approx 0.3$ is the average terrestrial albedo, $\omega_i$ and $\phi_i$ with $i=1,2,3,\ldots$ are the planetary frequencies and phases, and $a_i$ are the amplitudes of these cyclical fluctuations.

We observe that a significant, although small decadal modulation, of the low cloud cover since 1980, appears synchronized to the decadal solar cycle \citep{Harrison,Svensmark2007,Brown}.  Also a careful analysis of a critical study \citep{Sloan2008} would approximately support the theory: their figure 1  shows the low cloud cover anomalies in the polar, intermediate and equatorial regions from 1980 to 2005, and a sufficiently clear decadal cycle, in synchrony with the decadal solar cycle, is observed.
Sloan and Wolfendale's figure 1 suggests that the decadal cycle in cloud cover has amplitude of about 2\%, and it is in synchrony with the decadal solar cycle. Moreover, the same figure shows a downward multidecadal trend in cloud cover of about 4\% from 1980 to 2000 that could well agree with the 60-year modulation of the auroras and in the global temperature observed above in Figure 2. The percentages relative to the decadal and 60-year modulation would approximately correspond to about 0.5\% and 1\% change of the Earth's albedo respectively because these clouds may cover about 30-40\% of the Earth. Indeed, \cite{Wild} showed that a global solar brightening occurred from 1910 to 1940 and from 1970 to 2000, and   global solar dimming occurred from 1940 to 1970: see his figure 9. This 60-year reversal from solar brightening to dimming and from solar dimming to brightening was likely due to a cloud cover 60-year modulation and would support the theory herein presented.

\begin{figure}[t!]
\includegraphics[angle=0,width=21pc]{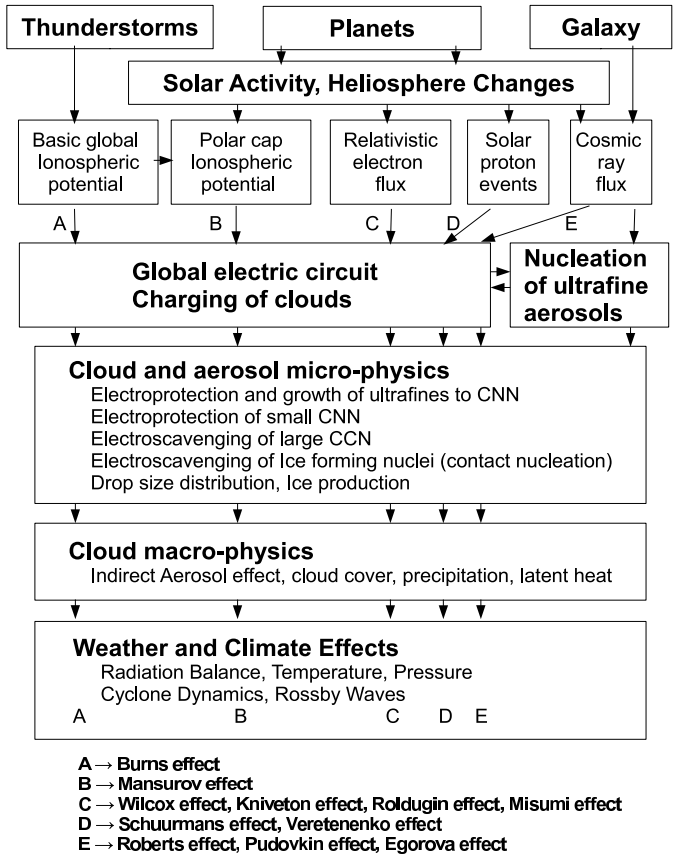}
\caption{ Connections between weather and climate with thunderstorms, solar activity, and galactic cosmic ray flux, via the global atmospheric electric circuit and cloud and aerosol microphysics. Solar and heliospheric activity is supposed to be partially driven by the gravitational and magnetic fields of the planets, as they move around the Sun. See Tinsley (2008) for more details about the single mechanisms reported in the figure. }
\end{figure}

An about 0.5\% and 1\% albedo change would greatly amplify the climate sensitivity to total solar irradiance, as shown in the above exercise and could also be quantitatively compatible with a 0.1 $^oC$ climate signature associated to the decadal astronomical cycle \citep{Scafetta2009} and with a 0.3/0.4 $^oC$ natural warming observed from 1970 to 2000, as deduced from the 60-year cyclical modulation of the temperature depicted in Figure 2. These results would be compatible with the empirical climate sensitivity estimated by \cite{Scafetta2009}. Thus, the observations appear to be consistent with the theory of an oscillating albedo.

Figure 9 summarizes a chain of mechanisms that, according to the finding of this paper, reasonably links the planetary motion to climate change through solar and heliosphere modulation of the magnetosphere and ionosphere that regulate the cloud cover percentage. The figure updates a figure from \cite{Tinsley} where more details about the single physical mechanisms are found. The planets, in particular Jupiter and Saturn, may modulate solar activity, the heliosphere and the magnetosphere in proximity of the Earth.  Therefore, a cyclical modulation of the electric properties of the ionosphere occurs, as revealed by the aurora record. This modulation causes a synchronized change of the cloud system through the atmospheric electric circuit. The modulation of the cloud system, together with the solar variations, drives all other climate subsystems including the ocean. All subsystems of the climate synchronize \citep{Scafetta2010b,Strogatz}. The resulting output is a global climate system that oscillates with the planetary frequencies. An additional modulation with a period of about 9.1 years is caused by the solar/lunar tidal forces that would act more directly in the ocean and in minor degree on the magnetosphere and ionosphere.

In addition, the variation of the length of the day (LOD) of the Earth has been proposed as another mechanism linking planetary motion to climate change. LOD presents a 60-year cycle \citep{Stephenson,Klyashtorin2001,Roberts,Mazzarella,Mazzarella1,Morner,Scafetta2010b}.  Because the LOD 60-year modulation is negatively correlated to the 60-year climate cycle the following mechanism may be working. When the low cloud cover decreases there is an increase of the energy stored in the oceans that causes variations in atmospheric zonal wind circulation and eventually causes more water to move from the equator to the poles, causing a global warming. This toward-pole mass movement induces a LOD decrease because the Earth, by becoming a little bit more spherical, speeds up because of the decrease of its rotational inertia. The latitudinal distribution and transport of energy and momentum associated to changes of zonal winds induced by cloud cover solar-induced changes have been recently advocated also for explaining the presence of a significant solar signature in the semiannual LOD variation \citep{Courtillot117}.

\section{Reconstruction and forecast of the climate oscillations based on aurora cycles}

According Eq. \ref{yu67},  if we can assume that solar irradiance and/or the albedo periodically oscillate, then the temperature at the surface would also oscillate with a similar set of frequencies. In fact, in first approximation, the surface temperature variation would be proportional to the incoming irradiance $\Delta F$.  However, the same result would be obtained also if the heat capacity of the Earth is taken into account. For example, a simple energy balance model could be given by a first order differential equation of the type:

\begin{equation}\label{iuytr}
    \frac{d\Delta T(t)}{dt} = \frac{k \Delta F(t)-\Delta  T(t)}{\tau},
\end{equation}
 where $k$ is a climate sensitivity to the forcing and $\tau$ is the thermal relaxation  time of the system \citep{Scafetta2009}. As well known, if the forcing is of the type $\Delta F(t)=\cos(\omega t)$, the solution of the above equation is
\begin{equation}\label{iuytr2}
   \Delta T(t)=k \frac{\cos( \omega t) +\omega \tau \sin(\omega t)}{1+\omega^2 \tau^2} +c = A \cos(\omega t + \phi) +c
\end{equation}
where $\phi$ and $A$ are  given phase and amplitude that depend on $\omega$ and $\tau$,  and $c$ is a given constant. Indeed, this result emerges from  quite general mathematical properties  under the approximation of small forced oscillations, and would be also obtained with professional climate GCMs. Thus, if the solar irradiance and/or the albedo are oscillating with a given set of frequencies, then the surface temperature too would present a harmonic component made of the same set of frequencies (in general there may be also sub- and super-harmonics). We would have:

\begin{equation}\label{iuytr33}
  \Delta T(t)= A_0 + \sum A_i  \cos(\omega_i t + \phi_i).
\end{equation}
Therefore, as already noted by Lord Kelvin to solve the tidal problem \citep{Thomson}, if  the purpose is only to determine the response of the climate system to a harmonic forcing,  general mathematical theorems allow us to bypass the problem of accurately solving the single physical mechanisms. Indeed,  the parameters $A_i$ and $\phi_i$, which would contain all physical information about the climate system, could be easily measured and determined by simple regression on the actual temperature observations.

It is necessary to test whether a harmonic model based on aurora cycles can efficiently both reconstruct and forecast the climate oscillations. In fact, a regression model that uses a set of parameters would be able to fit only the region used to calibrate the model, but if the model is not made of the right dynamical components it will fail any forecasting.

We proceed by  assuming, in first approximation, that the global surface temperature record has a harmonic component  characterized by a basic set of five periodic signals with periods equal to 9.1, 10.5, 20, 30 and 60 years, as  found in the aurora records and in the planetary motion \citep{Scafetta2010b}.
Because the temperature rose from 1850 to 2010, this upward secular trend needs to be detrended from the data because the proposed harmonic model is limited to decadal and multidecadal cycles. We use two fit functions of the type:

\begin{equation}\label{F1}
    F_1(t) = A_1 (t-1850)^2 + B_1
\end{equation}
and

\begin{equation}\label{F2}
    F_2(t) = A_2  (t-1850)^2 + B_2~.
\end{equation}
We use $F_1(t)$ to fit the temperature record  from 1850 to 2010. We use $F_2(t)$ to fit the temperature record from 1850 to 1950. We find that $A_1 = 2.9\cdot10^{-5} \pm 2\cdot10^{-6} ~^oC/y^2$ , $B_1= -0.41 \pm 0.02 ~^oC$, and that $A_2= 2.7\cdot10^{-5} \pm 2\cdot10^{-6}~^oC/y^2$, $B_2= -0.40 \pm 0.02~^oC$. We observe that the two acceleration coefficients $a_1$ and $a_2$ do not drastically change. This fact may indicate that in 1950 a quadratic fit of the temperature from 1850 to 1950 would have well forecasted the observed 1950-2010 temperature upward trend, but this result is likely a coincidence.

Indeed, the secular upward warming trend  may be due to a combination of a millenarian cycle, of a bi-secular cycle,  of  poorly corrected urban heat island effects \citep{McKitrick2007,McKitrick2010} and of anthropogenic GHG emissions. \cite{Scafetta2009,Scafetta2010a} and \cite{Loehle11} showed that at least 50\% of the upward secular trend of the temperature since 1900 could be induced by the correspondent secular increase of the solar activity during this same period. Here we do not investigate further this issue because the above quadratic polynomial fit is not part of the astronomical model, but only a convenient way to represent the trending of the temperature from 1850 to 2010, not outside this period.

\begin{figure}[t!]
\includegraphics[angle=0,width=21pc]{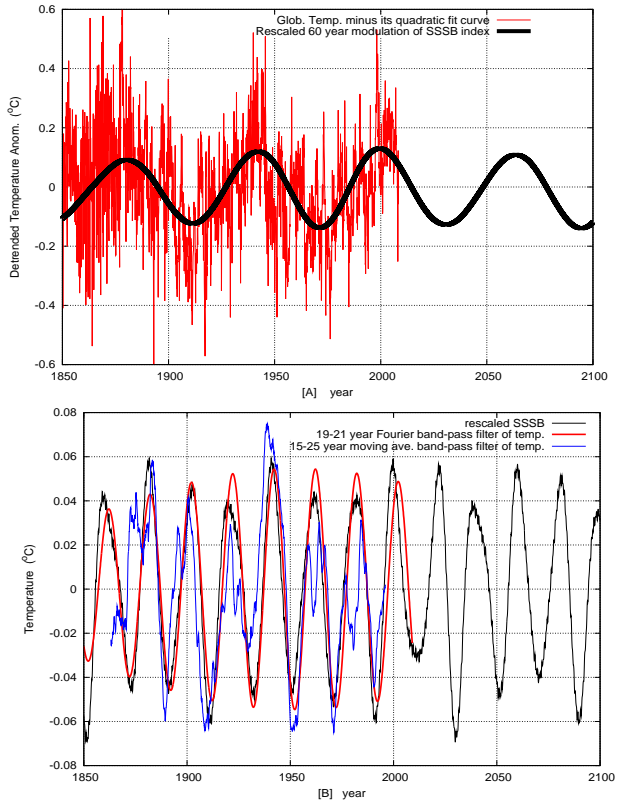}
\caption{ [A] Rescaled  60-year modulation of the solar speed relative to the solar system barycenter (SSSB) (black) (see Figure 6B) against the global surface temperature record  detrended of its quadratic fit. [B] Rescaled modulation of the solar speed relative to the solar system barycenter (black) (see Figure 6B) against two alternative pass-band filtered records of the temperature around its two decadal oscillations. The figures clearly indicate a strong coherence between the astronomical oscillations and the oscillations observed in the climate system.  }
\end{figure}

That the harmonic model may likely work is already implicit in Figure 10. Here, it is observed a very good synchrony between the quasi 20 and 60-year oscillations of the solar system, which are mostly driven by Jupiter and Saturn, and the correspondent oscillations observed in the climate system.

The astronomical harmonic model must be tested on its forecasting capabilities to be credible. We consider two independent time intervals (1850-1950 and 1950-2010):

\begin{eqnarray*}
  P_1(t) &=& a_1 \cos\left(\frac{2\pi(t-T_1)}{60}\right)+b_1 \cos\left(\frac{2\pi t}{30}+\alpha_1\right)+ \\
   & &  c_1 \cos\left(\frac{2\pi(t-T_1)}{20}\right) + d_1 \cos\left(\frac{2\pi t}{10.5}+\beta_1\right) +\\
   & &  e_1 \cos\left(\frac{2\pi t}{9.1}+\gamma_1\right)+ f_1
\end{eqnarray*}
and

\begin{eqnarray*}
  P_2(t) &=& a_2 \cos\left(\frac{2\pi(t-T_2)}{60}\right)+b_2 \cos\left(\frac{2\pi t}{30}+\alpha_2\right)+ \\
   & & c_2 \cos\left(\frac{2\pi(t-T_2)}{20}\right)+ d_2 \cos\left(\frac{2\pi t}{10.5}+\beta_2\right) +\\
   & & e_2 \cos\left(\frac{2\pi t}{9.1}+\gamma_2\right)+f_2\\
\end{eqnarray*}
In our simplified model, we use $T_1=2000.5$ and $T_2=1941$, which correspond to the dates of the last two J/S conjunctions of Group I, see Table 2. This choice is also supported by the finding depicted in Figure 10.

We proceed by analyzing three periods. We use the function $F_1(t)+P_1(t)$ to fit the temperature data during two periods:  from 1850 to 2010 and from 1950 to 2010.  We use the function $F_2(t)+P_2(t)$ to fit the temperature data   from 1850 to 1950.   The coefficients of the three regression models are reported in Table 3. We find that the three sets of regression coefficients coincide within the error of measure: both the amplitudes and the phases coincide. The reduced $\chi_o^2$ is $0.34$ with 9 degrees of freedom and the coherence probability between the two models is $P_9(\chi^2 \geq \chi^2_o)>95\%$. Thus, within the limits of the data herein used, it is found that the astronomical harmonic model can both reconstruct and forecast the temperature signal within a high statistical confidence level.

\begin{figure}[t!]
\includegraphics[angle=0,width=21pc]{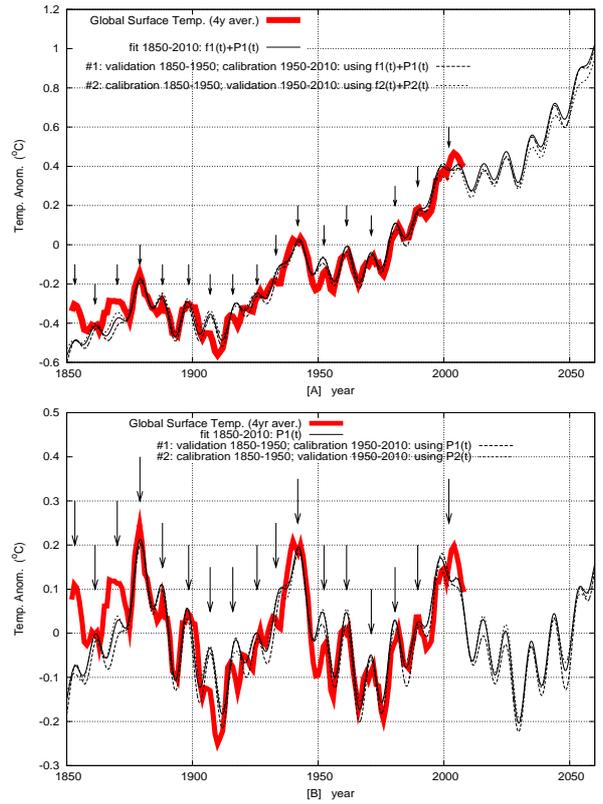}
\caption{ Astronomical harmonic constituent model reconstruction and forecast of the global surface temperature. [A] Four year moving average of the global surface temperature against the climate reconstructions obtained by using the function $F_1(t) + P_1(t)$ to fit the period 1850-2010 (black solid) and  the period 1950-2010 (dash), and the function $F_2(t) + P_2(t)$ to fit the period 1850-1950 (dots). [B] The functions $P_1(t)$ and $P_2(t)$  represent the periodic modulation of the temperature reproduced by the celestial model based on the five aurora major decadal and multidecadal frequencies. The arrows indicate the local decadal maxima where the good matching between the data patterns and the models is observed. Note that in both figures the three model curves almost coincide for more than 200 years and well reconstruct and forecast the temperature oscillations. }
\end{figure}

Figure 11A depicts the three regression model outputs against the global surface temperature, which has been smoothed with a four-year moving average algorithm. Figure 11B depicts only the oscillating functions $P_1(t)$, which fits  the periods 1850-2010 (solid) and 1950-2010 (dash), and the function $P_2(t)$, which  fits  the period 1850-1950 (dots). Figure 11A and 11B clearly show a very good matching during the entire period 1850-2010 between the decadal and multidecadal climate oscillations and the oscillations reconstructed by the models in all three cases.

The result depicted in the figure implies that if we were in 1950, by using an astronomical harmonic constituent model based on aurora cycles and on the temperature record from 1850 to 1950, it could have been possible to forecast with a reasonable accuracy the climate oscillations within a decadal and multidecadal scale occurring from 1950 to 2010.  The figure also implies the opposite, that is, that the  decadal and multidecadal climate oscillations from 1850 to 1950 could be forecasted by using the same astronomical harmonic constituent  model calibrated on the temperature data from 1950 to 2010. Note that during the period 1915-1940 the decadal cycle appear to disappear because during this period the 9.1 and 10.5 year cycles interfere destructively.
The remarkable result depicted in Figure 11 does imply that the climate oscillations within the secular scale can be accurately forecasted by using the major decadal frequencies found in the mid-latitude aurora records, which appear to be driven by astronomical cycles associated to the movement of Jupiter and Saturn (about 10, 20, 30 and 60 years), to the lunar cycle (9.1 years) and to the 11-year solar cycle.

About the imminent future, Figure 11A suggests that if the quadratic background upward trend of the temperature continues during the next decades, the global temperature may remain approximately constant until 2030-2040, and this result would be approximately compatible with that found in \cite{Loehle11} by using a different methodology that yield a net anthropogenic warming component of about +0.66 $^oC/century$ from 1940 to 2050 against the IPCC estimate of +2.3 $^oC/century$ during the same period.
 However, the quadratic trend used here should not be trusted after 2040 also because is not part of the harmonic model, but only a convenient way to fit the temperature from 1850 to 2010.

 The observed warming since 1850 could have been partially caused by a large quasi-millennial climate cycle, which has caused the Medieval Warm Period, the Little Ice Age and the Modern Warm Period. Probably this large quasi-millennial cycle just entered or is going to enter into its cooling phase and, therefore, it can further cool the planet on a multi-secular scale.
Other cooling natural component can derive from a bi-secular natural solar induced cycle. These cycles are known to have regulated multidecadal cold periods (the well-known Oort, Wolf, Sp\"orer, Maunder, Dalton minima). In fact, as noted in \cite{Scafetta2010a} the last four sunspot cycles (\#20-21-22-23) look quite similar to the four sunspot cycles (\#1-2-3-4) that occurred just before the Dalton minimum. That the Sun may be entering in a prolonged period of minimum activity is reasonable \citep{Duhau}.
That a multi-secular climate natural cycle may be turning down is also indirectly suggested by the fact that since 1930 the sea-level acceleration has been, on average, slightly negative \citep{Houston}. Interestingly, the measured deceleration of the sea-level rise from 1930 to 2010 occurred despite the strong positive acceleration of anthropogenic GHG emissions during the same period, which would indicate that other more powerful forcings (such as an astronomical forcing of the cloud system) are the major regulators of climate change.

\section{Conclusion}

Four centuries ago, Johannes Kepler explained that earthly nature couldn't help but respond to the dictates of heavenly harmonies, and said that nature is affected by an aspect \emph{just as a farmer is moved by music
to dance} \citep{Kemp}. Kepler clearly shared the common belief of his time that the climate was influenced by astronomical cycles and understood the subtle phenomenon of \emph{collective synchronization} that has been extensively studied in nonlinear complex science since the times of Huygens \citep{Pikovsky,Strogatz,Scafetta2010b}.
Indeed, climate change records present geometrical characteristics that suggest that climate is synchronized, probably through the Sun and the heliosphere, to complex astronomical cycles driven by planetary harmonics.

In this paper, we have studied the historical record of mid-latitude auroras from 1700 to 1900, and of the Faroes Islands from 1873 to 1966 to search for a possible physical mechanism linking planetary motion to climate. We have shown that mid-latitude aurora records and the global surface temperature record share a set of oscillations with periods of about 9.1, 10-11, 20-21, 30 and 60-62 years. Other shorter and longer oscillations may be present, but they are not discussed here. In particular, clear quasi 60-year cycles in the aurora records are synchronized to the 60-year cycle observed in the global surface temperature and in multi-secular proxy climate reconstructions of both the Atlantic and Pacific climatic oscillations since 1650 and in Indian summer monsoon records. \cite{Charvatova1988} found that a large 60-year cycle is present in the mid-latitude aurora record also for the longer period from 1001 to 1900 and other planetary frequencies are present as well in the millennial aurora record. Numerous other studies have found quasi 10, 20 and 60-year cycles in multiple climate records and in astronomical records, as summarized in the Introduction.  By taking into account the results of \cite{Scafetta2010b}, this synchrony exists also with the global ocean and land global surface temperature records of both hemispheres.

The aurora record cycles reveal a direct or indirect planetary influence on the Sun and on the Earth's magnetosphere and ionosphere.
Indeed, proxy reconstructions may suggest a 60-year cycle in the total solar irradiance (TSI) was almost stable or slightly decreased from 1880 to 1910, and it increased from 1910 to 1940. From 1940 to 1970 TSI may have decreased as the TSI reconstruction of \cite{Hoyt} suggests. Hoyt and Schatten's TSI reconstruction well correlates with the temperature records during the last century \citep{Soon}. Finally, an increase of the solar activity from 1970 to 2000 and a decrease afterward would be supported by the ACRIM TSI satellite composite, which may more faithfully reproduce the satellite observations \cite{Willson,ScafettaW2009,Scafetta20112}  (but not by the PMOD composite \cite{Frohlich}, which is the TSI record preferred  by the \cite{IPCC}). Thus, solar activity could have been characterized by a quasi 60-year modulation superimposed to other larger secular, bi-secular and millennial cycles \citep{Krivsky1984,Ogurtsov}, although it may not appear evident in every total solar irradiance reconstruction \citep{Krivova,Wang}.

It is possible that when Jupiter and Saturn are closer to the Sun, there may be an increased solar activity because of the stronger planetary tides and other mechanisms \citep{Wolff}, and a stronger magnetic field within the inner region of the solar system form, although the patterns may be more complicated because of the presence of other cycles that will be discussed in another paper.  A stronger solar or heliospheric magnetic field better screens galactic cosmic ray fluxes. Fewer cosmic rays reaching the Earth imply a weaker ionization of the upper atmosphere. As a side-effect, less auroras form in the middle latitudes because a stronger magnetic field and a less ionized ionosphere mostly constrains the auroras in the polar region. In addition, the level of ionization of the atmosphere has been proposed as an important mechanism that can modulate the low cloud cover formation \citep{Tinsley,Kirkby,Svensmark2009}. Essentially, when the ionization is weaker, less clouds form. A solar and heliospheric modulation of the cloud system would greatly contribute to climate change through an albedo modulation (see Eq. \ref{albedo}). The above sequence of mechanisms would explain why the climate presents oscillations at multiple frequencies that are synchronized with the aurora and the planetary cycles.

We have used a phenomenological harmonic model based on five decadal and multidecadal frequencies with periods of 9.1, 10.5, 20, 30 and 60 years that has been detected in the aurora records and that could be associated to evident astronomical and solar/lunar tidal cycles. We have shown that it is possible to forecast the climate oscillations occurred from 1950 to 2010 using the climatic information derived from the period 1850-1950 and the frequency information deduced from the mid-latitude aurora before 1900. Analogously, we have shown that it is possible to forecast the climate oscillations that occurred backward from 1850 to 1950 using the information derived from the period 1950-2010. Thus, these findings strongly support the thesis that the climate oscillations can be approximately forecasted by using astronomical cycles. The proposed astronomical harmonic constituent model for climate change based on aurora cycles is conceptually equivalent to the commonly used tide-predicting machines based on planetary harmonic constituent analysis conceived by Lord Kelvin in 1867.

 Interestingly, the traditional Chinese, Tamil and Tibetan calendars are arranged in major 60-year cycles \citep{Aslaksen}.  Perhaps, these sexagenarian cyclical calendars were inspired by climatic and astronomical observations and were used for timing and regulating human business. For example, even ancient Sanskrit texts report about a 60-year monsoon rainfall cycle \citep{Iyengar} and associate it to the movement of Jupiter and Saturn, the \emph{ Brihaspati} 60-year cycle, which may explain why Asian populations used sexagesimal calendars. Indeed, a 60-year cycle linked to Jupiter and Saturn was extremely well known to several ancient civilizations \citep{Temple}.   The major cycles discussed in this paper are also approximately found in the major business cycles \citep{Pustilnik,Korotayev}. A link between planetary motion, climate and economy (which mostly in the past could be driven by agricultural productivity) would ultimately explain the interest of the ancient civilizations in tracking the position of the planets and their attempts in developing multiscale cyclical calendars \citep{Ptolemy,Masar,Swerdlow}.

 In conclusion, the results presented here strongly support and reinforce the argument of \cite{Scafetta2010a,Scafetta2010b} that the climate is forced by astronomical  oscillations related to the Sun, the Moon and the planets, and, as Figure 11 shows, a significant component of it can be forecasted within an acceptable uncertainty with appropriate harmonic models.

\newpage

\onecolumn

\begin{table}
  \centering
\begin{tabular}{|c|c|c|c|c|c|}
  \hline
Cycle	&	Temp/Fig4A	&	AAR	&	fig4B	&	fig4C	&	fig4D	\\ \hline
\#1	&	$9.1 \pm 0.2$	&	$9.15 \pm 0.2$	&	$9.1 \pm 0.2$	&	$9.2 \pm 0.2$	&	n/a	\\ \hline
\#2	&	$10.4 \pm 0.3$	&	$10.33 \pm 0.3$	&	$10.0 \pm 0.3$	&	$10.2 \pm 0.3$	&	$10.8 \pm 0.3$	\\ \hline
\#3	&	$20.9 \pm 0.9$	&	$20.6 \pm 0.9$	&	$21.3 \pm 0.9$	&	$20.3 \pm 0.9$	&	$20.2 \pm 0.9$	\\ \hline
\#4	&	$32 \pm 2$	&	$29.5 \pm 2$	&	$29 \pm 2$	&	$30 \pm 2$	&	n/a	\\ \hline
\#5	&	$62 \pm 5$	&	$62 \pm 5$	&	$62 \pm 5$	&	$64 \pm 8$	&	$60 \pm 5$	\\ \hline
\end{tabular}
  \caption{First column: periods of the cycles \#1, \#2, \#3, \#4 and \#5 found in the global surface temperature: see Figure 4A. Third column:  periods of the cycles \#1, \#2, \#3, \#4 and \#5 found in the 1700-1966 aurora record: see Figure 4B. Fourth column:  periods of the cycles \#1, \#2, \#3, \#4 and \#5 found in the 1700-1880 aurora record: see Figure 4C. Fifth column:  periods of the cycles \#2, \#3 and \#5 found in the 1872-1966 aurora record: see Figure 4C. Second column: average of the periods of the cycles \#1, \#2, \#3, \#4 and \#5 found in the three aurora records. The results depicted in the first and second column are perfectly coherent within the error of measure for each couple of values. For the 5 frequency couples combined, the reduced $\chi^2$ is $\chi^2_o=0.18$ with 5 degrees of freedom ($P_5(\chi^2 \geq \chi^2_o)>96\%$).  }\label{}
\end{table}

\begin{table}
  \centering
\begin{tabular}{|c|c|c|c|c|c|c|c|}
  \hline
	&	Jupiter	&		&		&	Saturn	&		&		&	tidal	\\
Group I	&	Right asc.	&	Decl.	&	Dist	&	Right asc.	&	Decl.	&	Dist	&	elong.	\\ \hline
Jun 23, 2000	&	3h 19m 29.7s	&	17$^o$  19' 48''	&	4.996	&	3h 20m 46.9s	&	16$^o$  09' 35''	 &	 9.148	&	 1801	 \\ \hline
Nov 15, 1940	&	2h 38m 22.5s	&	14$^o$  17' 03''	&	4.975	&	2h 40m 02.0s	&	13$^o$  06' 15''	 &	 9.206	&	 1822	 \\ \hline
Apr 3, 1881	&	1h 59m 49.6s	&	10$^o$  56' 01''	&	4.960	&	2h 01m 41.6s	&	9$^o$  46' 24''	 &	 9.271	 &	 1836	 \\ \hline
Sep 16, 1821	&	1h 23m 54.9s	&	7$^o$  27' 24''	&	4.951	&	1h 25m 41.6s	&	6$^o$  19' 26''	 &	 9.339	 &	 1843	 \\ \hline
Feb 23, 1762	&	0h 50m 01.2s	&	3$^o$  56' 49''	&	4.948	&	0h 51m 51.2s	&	2$^o$  53' 04''	 &	 9.408	 &	 1846	 \\ \hline
Aug 8, 1702	&	0h 17m 53.7s	&	0$^o$  30' 16''	&	4.949	&	0h 19m 43.3s	&	-0$^o$  28' 12''	 &	 9.477	 &	 1843	 \\ \hline
 $\sim$	&		&		&		&		&		 &		 &		\\ \hline
Mar 9, 1206	&	3h 38m 35.0s	&	18$^o$  45' 16''	&	5.054	&	3h 39m 46.7s	&	17$^o$  39' 14''	 &	 9.027	 &	1747	 \\ \hline
Jul 26, 1146	&	2h 55m 26.0s	&	15$^o$  47' 32''	&	5.023	&	2h 56m 58.1s	&	14$^o$  38' 17''	 &	 9.064	&	 1776	 \\ \hline
Dec 14, 1086	&	2h 14m 02.8s	&	12$^o$ 20' 48''	&	4.998	&	2h 15m 45.0s	&	11$^o$ 10' 10''	 &	 9.116	 &	 1800	 \\ \hline
Group II	&		&		&		&		&		&		&		\\ \hline
Apr 17, 1981	&	12h 28m 16.4s	&	-1$^o$  38' 06''	&	5.451	&	12h 29m 57.6s	&	-0$^o$  38' 21''	 &	 9.563	&	 1395	 \\ \hline
Aug 23, 1921	&	11h 51m 01.8s	&	2$^o$  21' 51''	&	5.442	&	11h 52m 34.2s	&	3$^o$  14' 38''	 &	 9.476	 &	 1404	 \\ \hline
Dec 28, 1861	&	11h 13m 17.8s	&	6$^o$  20' 14''	&	5.427	&	11h 14m 40.1s	&	7$^o$  04' 21''	 &	 9.393	 &	 1418	 \\ \hline
May 8, 1802	&	10h 35m 26.9s	&	10$^o$  06' 06''	&	5.405	&	10h 36m 27.1s	&	10$^o$  41' 21''	 &	 9.314	 &	1436	 \\ \hline
Sep 18, 1742	&	9h 57m 07.1s	&	13$^o$  33' 53''	&	5.378	&	9h 57m 49.9s	&	13$^o$  58' 48''	 &	 9.241	&	 1458	 \\ \hline
Feb 2, 1683	&	9h 18m 23.2s	&	16$^o$  36' 37''	&	5.347	&	9h 18m 43.6s	&	16$^o$  50' 56''	 &	 9.177	 &	1484	 \\ \hline
Group III	&		&		&		&		&		&		&		\\ \hline
Apr 16, 1960	&	19h 42m 18.9s	&	-21$^o$  40' 29''	&	5.146	&	19h 42m 10.4s	&	-21$^o$  23' 03''	 &	 10.028	&	 1635	 \\ \hline
Sep 28, 1901	&	19h 07m 48.3s	&	-22$^o$  40' 42''	&	5.177	&	19h 07m 38.6s	&	-22$^o$  13' 36''	 &	 10.053	&	 1607	 \\ \hline
Mar 12, 1842	&	18h 33m 28.3s	&	-23$^o$  12' 38''	&	5.207	&	18h 33m 23.7s	&	-22$^o$  36' 25''	 &	 10.072	&	 1580	 \\ \hline
Aug 26, 1782	&	17h 59m 51.0s	&	-23$^o$  16' 44''	&	5.236	&	17h 59m 50.7s	&	-22$^o$  32' 18''	 &	 10.082	&	 1555	 \\ \hline
Feb 11, 1723	&	17h 26m 46.3s	&	-22$^o$  54' 28''	&	5.264	&	17h 27m 03.5s	&	-22$^o$  02' 54''	 &	 10.086	&	 1531	 \\ \hline
Jul 31, 1663	&	16h 54m 47.2s	&	-22$^o$  08' 09''	&	5.291	&	16h 55m 15.1s	&	-21$^o$  10' 25''	 &	 10.083	&	 1509	 \\ \hline
\end{tabular}
  \caption{Approximates dates of the J/S conjunctions relative to the Sun, which slightly differ from the conjunction dates as relative to the Earth. The coordinate of the planets are in Right Ascension and Declination in heliocentric equatorial coordinates which are relative to the equatorial plane of the Earth. The last column report the approximate value of the tidal elongation in Km (see Eq. \ref{eqte}) at the orbit of the Earth, that is, at 1 AU from the Sun. }\label{}
\end{table}

\begin{table}
  \centering
\begin{tabular}{|c|c|c|c|c|}
  \hline
	&	$P_1(t)$ (1850-2009)		&	$P_1(t)$ (1950-2009)		&	$P_2(t)$ (1850-1950)		\\ \hline
a	&	0.103	$\pm$ 0.011	&	0.109	$\pm$ 0.018	&	0.098	$\pm$ 0.015	\\ \hline
b	&	0.022	$\pm$ 0.011	&	0.031	$\pm$ 0.018	&	0.020	$\pm$ 0.015	\\ \hline
c	&	0.040	$\pm$ 0.011	&	0.043	$\pm$ 0.018	&	0.036	$\pm$ 0.015	\\ \hline
d	&	0.027	$\pm$ 0.011	&	0.030	$\pm$ 0.018	&	0.027	$\pm$ 0.015	\\ \hline
e	&	-0.053	$\pm$ 0.011	&	-0.055	$\pm$ 0.018	&	-0.050	$\pm$ 0.015	\\ \hline
f	&	-0.006	$\pm$ 0.008	&	-0.024	$\pm$ 0.013	&	-0.006	$\pm$ 0.010	\\ \hline
$\alpha$	&	2.08	$\pm$ 0.5	&	1.84	$\pm$ 0.6	&	2.41	$\pm$ 0.7	\\ \hline
$\beta$	&	1.43	$\pm$ 0.4	&	1.81	$\pm$ 0.6	&	1.14 $\pm$ 0.6	\\ \hline
$\gamma$	&	-0.32	$\pm$ 0.2	&	-0.47	$\pm$ 0.3	&	-0.26	$\pm$ 0.3	\\ \hline

\end{tabular}
  \caption{Multi nonlinear regression coefficients to fit of the global surface temperature record using the functions $P_1(t)$ and $P_2(t)$. $P_1(t)$ is used to fit  two periods:  1850-2009 and 1950-2009. $P_2(t)$ is used to fit the period 1850-1950. Note that the three sets of coefficients coincide within the errors of measure. Thus, there exists an accurate statistical coherence between the three models. For the 9 coefficient couples relative to the independent results reported in the last two columns the reduced $\chi^2$ is $\chi^2_o=0.34$ with 9 degrees of freedom ($P_9(\chi^2 \geq \chi^2_o)>95\%$).   }\label{}
\end{table}

\newpage

\end{document}